\documentstyle[epsbox,referee]{aa}

\begin{document}

   \thesaurus{03         
               (11.09.2;  
               11.09.4;  
               11.11.1;)}  

   \title{On the degree of dust extinction in major galaxy mergers 
with dusty starburst}


   \author{K. Bekki 
          \inst{1}
\& Y. Shioya \inst{2}          }

   \offprints{K. Bekki}

   \institute{1 Division of Theoretical Astrophysics, National Astronomical Observatory, Mitaka,
    Tokyo, 181-8588, Japan \\
   email address: bekki@th.nao.ac.jp \\
2 Astronomical Institute, Tohoku University, Sendai, 980-8578, Japan \\
   email address: shioya@astr.tohoku.ac.jp}

   \date{Received  April 28,  2000; accepted August 8, 2000}

   \maketitle
  \markboth{The degree of dust extinction}{} 

   \begin{abstract}

  We numerically investigate how fundamental properties of 
  the spectral energy distribution (SED) of
  a major gas-rich galaxy merger with dusty starburst
  are determined by the initial orbital configuration of the merger. 
  We found that an infrared luminous galaxy with  dusty starburst
  formed by a nearly retrograde-retrograde merger
  suffers the most remarkable dust extinction of 
  stellar light and consequently shows very red colours.
Considering that a  retrograde-retrograde merger 
does not produce strong and long tidal tails, 
this result  suggests that a luminous infrared
galaxy  without clear signs of interaction 
and merging shows very large internal dust extinction and redder colours. 
  These numerical results furthermore
imply that the morphology of a luminous infrared galaxy  can correlate with
  the degree of internal dust extinction and thus with the shape of the SED, principally because 
  the morphology of a merger also depends strongly
  on the initial orbital configuration of the merger.
  Based on these results, we discuss the origin
  of ultra-luminous infrared galaxies. 

              
   \end{abstract}

\section{Introduction}

  Ultra-luminous infrared galaxies (ULIRGs), which are ongoing major mergers
  between gas-rich spirals, are generally considered to 
  provide valuable clues  to 
the understanding  not only for the formation processes
  of elliptical galaxies and active galactic nuclei (AGN) 
but also for the nature
  of high redshift dusty starburst galaxies (Sanders et al. 1988; 
  Sanders \& Mirabel 1996; 
Meurer et al. 1999;
  Trentham et al.  1999).
  In particular, to investigate how and
  to what degree stellar light is heavily obscured by
  dust in low redshift ULIRGs may 
   lead us to clarify the origin even for high redshift faint sub-millimeter 
  sources recently discovered by 
 the Sub-millimeter Common-User Bolometer Array
 (SCUBA) (Holland et al. 1999) on the James Clerk Maxwell Telescope
 (Smail et al. 1997;
 Hughes et al. 1998; Smail et al. 1998; Barger et al. 1999; Lilly et al.  1999).
 Recently, Trentham et al. (1999) have found that the degree of internal dust extinction
 at optical and ultraviolet wavelength in the spectral energy distribution (SED) 
 of ULIRGs is very different between ULIRGs,
though the total number of the sample ULIRGs is very small (only 3). 
They furthermore showed that the magnitude of the drop of SEDs 
from optical to ultraviolet band 
 ranges from a factor of $\sim$ 3 in IRAS F22491-1808 
to a factor of $\sim$ 100 in
VII Zw 031. 
 The derived SEDs of ULIRGs from ultraviolet to near-infrared band 
 not only help us to understand the physical relationship between low 
redshift  ULIRGs and high redshift 
 faint SCUBA sources  (Trentham et al. 1999)
 but also provides valuable information on an evolutionary link
 between faint SCUBA sources with blue  $V-I$  colour ($\sim$ 1.5 mag) 
observed by    Smail et al. (1998) 
and Extremely Red Objects 
 (EROs)  first discovered  by Elston et al. (1988).
 It is, however, not clear at all what controls the degree of internal dust extinction
 in ULIRGs.

 The purpose of this paper is to investigate what determines the nature of SEDs 
 and the degree of internal dust extinction 
 in ULIRGs by performing numerical simulations that can 
 follow chemodynamical and photometric evolution of major galaxy mergers with
 dusty interstellar gas in an explicitly
 self-consistent manner.
 We here consider the  global dynamical evolution of galaxy mergers,
 which depends strongly on initial orbital configurations of mergers
(i.e., whether a merger is prograde-prograde one
or retrograde-retrograde one),
 is $one$  of primarily important factors  for the SEDs of ULIRGs.
Mihos
\& Hernquist (1996) have already investigated in detail
the $gas$ $dynamical$ $evolution$ of major mergers and 
demonstrated that internal structure of merger progenitor disks
(i.e., whether a disk has a  strong bulge or not) rather than
orbital configurations is a key factor for the merger star formation
history. However they did not discuss the nature of SEDs of ULIRGs.
Therefore our present results are  complementary  to their
important results in the sense that we can demonstrate
how the $SEDs$ of mergers can depend on initial physical
conditions of major mergers. 
 A remarkable difference between the present study and previous ones
that investigate SEDs of galaxies
(e.g., Mazzei et al. 1992; Franceschini et al.  1994;
Gordon et al.  1997; Guiderdoni et al. 1998) 
is that dynamical effects
on stellar and gaseous distributions are included in the present study. 
 We suggest  that the SED of an  ULIRG
depends on the orbital configuration
of a merger and furthermore that the SED of the  ULIRG can correlate strongly 
with its morphology.
Physical correlations between morphological properties and SEDs
in ULIRGs have not been discussed so extensively in previous studies. 
The present numerical results thus shed new light on
formation and evolution  of ULIRGs.

\section{Model}

We here investigate both temporal  evolution of galactic morphology
and that of SED,
based on  numerical simulations that can follow 
both  dynamical and chemical evolution of galaxies.
The numerical techniques for solving galactic 
chemodynamical and photometric evolution
and the methods for deriving SEDs in numerical
simulations of galaxy mergers with dusty starburst
are given in  Bekki \& Shioya (1999) and Bekki et al. (1999),
respectively.
Accordingly we describe only
briefly the present numerical model of dusty starburst galaxy mergers here. 
 The physical parameters used in this paper are very similar to
those adopted by Bekki et al. (1999) and plausible and realistic
for gas-rich major mergers.
 We construct  models of galaxy mergers between gas-rich 
 disk galaxies with equal mass by using Fall-Efstathiou model (1980).
 The total mass 
 and the size   of
a progenitor disk are 6.0 $\times$ $10^{10}$ $ \rm M_{\odot}$ 
 and 17.5 kpc,  
 respectively. 
  The collisional and dissipative nature 
  of the interstellar medium 
 with the initial gas mass fraction $ f_{\rm g}$ = 0.2
is  modeled by the sticky particle method
  (\cite{sch81}).
  Star formation 
  is modeled by converting  the collisional
  gas particles
  into  collisionless new stellar particles according to 
  the Schmidt law (Schmidt 1959)
  with the exponent of  2.0.
 As long as the exponent
ranges from 1.5 to 2.0,
our numerical results do not depend so strongly on the
exponent of the Schmidt  law.
 Chemical enrichment through star formation during galaxy merging
is assumed to proceed both locally and instantaneously in the present study.
The fraction of gas returned to interstellar medium in each stellar particle
and the chemical yield
are 0.3 and 0.02, respectively.
Initial metallicity $Z_{\ast}$ for each stellar and gaseous
particle in a given galactic radius  $R$  (kpc) from the center
of a disk is given 
according to the observed relation $Z_{\ast} = 0.06 \times {10}^{-0.197 \times (R/3.5)}$
of typical late-type disk galaxies (e.g., Zaritsky et al. 1994).

In the present study,
we consider that the most important parameter for determining morphological
and photometric evolution of a galaxy merger
is 
an initial orbital configuration of the  merger.
We accordingly investigate merger models with  variously  different 
orbital configurations. 
We here present the results of two models, a nearly prograde-prograde
model (referred to as $PP$ model for convenience),  
and a nearly retrograde-retrograde one ($RR$).
The reason for this adoption is that these models
clearly show the most remarkable differences in the morphological and
photometric evolution of galaxy mergers between these two. 
The results for models with variously different orbital configurations
will be  described in detail by Bekki \& Shioya (2000).
The angle between intrinsic spin vector 
and orbital one in one disk for  each of mergers
is set to be 30 degrees.
 The orbital plane of a  galaxy merger is assumed to be the same as $x$-$y$ plane
 and the initial distance between the center of mass of merger progenitor
 disks is 140 kpc.
 Two disks in the merger are assumed to encounter each other parabolically
 with the pericentric distance of 17.5 kpc.

Structural and kinematical properties of merger remnants,
   star formation history of mergers with prominent bulges,
   and that of dusty starburst mergers between bulgeless spirals
   have been already given in Barnes \& Hernquist (1992),
   Mihos \& Hernquist (1996), and Bekki et al. (1999), respectively.
   Thus we describe only morphological and photometric properties
   of mergers between two gas-rich bulgeless disk galaxies
   $at$ $the$ $epoch$ $of$ $massive$ $starburst$ in the present study.  
For calculating the SED
of a merger, 
we use the 
spectral library GISSEL96 which is the  latest version of  Bruzual \& Charlot (1993).
 Using  the derived SED, 
 we also investigate how dust-enshrouded starburst galaxy mergers
 at z=0.4 and  1.0 can be seen  with
 the  $Hubble$ $Space$ $Telescope$ ($HST$). 
 The method to construct the  synthesized $HST$ images of galactic morphology
 in the present study is basically the same as that described by Mihos (1995).
 For calculating the SED of a merger at z=0.4 and  1.0,
 we assume that 
 mean ages of old stellar components
 initially in  a merger progenitor  disk at the redshift z=0.4 and  1.0 are
 7.14 and  3.80 Gyr, respectively.
 The $V$ band apparent  magnitude $m_{V}$, global colours
($B-V$,  $V-R$,  $V-I$,  $R-K$, and   $I-K$), and  $S_{\rm 850 \mu m}$ flux
at each redshift (z=0.4, 1, 1.5, 2, and 3) for the two models ($PP$ and $RR$)
are summarized in the Table 1.
 In the followings, the cosmological parameters
 $\rm H_0$ and $\rm q_0$ are  set to be 50 km s$^{-1}$ Mpc$^{-1}$ and 0.5
 respectively.

\section{Results}

Fig.1 shows star formation history both for  the $PP$ 
and $RR$ models. Although the strength of the maximum starburst
is not  very  different between the two models, the epoch
of the starburst is very different.
The massive starburst is triggered 
in the first encounter for the $PP$ model whereas
the starburst occurs only in the late merger phase
when the two cores merge to form a single (elliptical) galaxy
for the $RR$ model. Accordingly the separation of two galactic cores
at the starburst epoch is very different (This is described in detail later
in Fig.3 and Fig.4).
The essential reason for 
this difference is that strong stellar bars, which can drive a large
amount of interstellar gas into the central region and thus trigger
massive starburst, can be formed in the first encounter $only$ for
the $PP$ model.
This result that orbital configuration of major galaxy merging
is an important determinant for merger star formation history
is in striking contrast to the previous numerical results by Mihos
\& Hernquist (1996).
They demonstrated that internal structure of merger progenitor disks
(i.e., whether a disk has a remarkable bulge or not) rather than
orbital configurations is a key factor for the merger star formation
history.
The reason for this  apparent difference between the present results and those
by Mihos \& Hernquist (1996) is essentially that
we do not include bulges at all in the present merger simulations.
As  pointed out  by  Mihos \& Hernquist (1996),
irrespectively of merger orbital configurations, 
massive starburst 
occurs in  the late merger phase when two cores
merge to form an elliptical galaxy for mergers with remarkable
bulges (with the mass ratio of bulge to disk $\sim$ 0.3:1),
primarily because bulge can greatly suppress the formation
of stellar bars during galaxy merging.
Accordingly we stress that the derived clear difference in star formation
history between the two models (i.e., the $PP$ and $RR$ models)
is true only for mergers between two disks without remarkable bulges.

Fig.2 describes the morphology of a merger at the epoch
of its maximum  starburst for 
the $PP$ and $RR$ models.
A merger in the present model is the most heavily obscured by dust
and shows the strongest far-infrared and sub-millimeter flux 
at this  starburst epoch. 
As is shown in Fig.2, 
young stellar components formed during merging 
are more compactly distributed in a merger than 
old stellar ones 
initially located within disks 
both for the $PP$ model and for the $RR$ one. 
This result indicates that 
compact young stellar components  are more heavily obscured by
dust than diffuse old ones in a gas-rich major merger. 
For the $PP$ model, the first and the strongest starburst is triggered
at $T$=0.7 $\sim$ 0.8 Gyr, 
principally because non-axisymmetric stellar bars formed in the first
encounter of the merger excite efficient gas transfer toward the
nuclear region of the two disks.
Consequently the morphology of the $PP$ merger at $T$=0.7 Gyr shows two
strong and long tidal arms and central bars.
As is shown in Fig.3, dusty interstellar gas is so efficiently transferred
to the central region of the stellar bar that the gas can
obscure the central starburst component in this $PP$ model. 
On the other hand, for the $RR$ model,
the maximum starburst occurs when two disks 
finally merge to form an elliptical galaxy ($T$=1.6 $\sim$ 1.7 Gyr).
Strong and long tidal tails are not 
developed during merging in this $RR$ model,
accordingly the morphology at $T$=1.6 Gyr appears to show no
clear signs of tidal interaction.
As is shown in Fig.4, two starburst cores composed mainly of
very young stars can be clearly seen in the central 3.5 kpc
of the $RR$ merger.
The old stellar components, on the other hand, 
are  more diffusely distributed and do not show
such distinct two cores in the central region of the $RR$ merger.
Dusty interstellar gas is distributed such that
it surrounds  the two starburst cores.
This mass distribution of gas and stars is similar to
that found in  Arp 220 (e.g., Norris 1985; Sakamoto et al. 1999),
 which is  an ULIRG without long tidal tails
(e.g., Sanders \& Mirabel 1996; Ohyama et al. 1999).
This result for the $RR$ model accordingly suggests that 
Arp 220 is formed by a nearly retrograde-retrograde merger.
These results for the above two models
clearly demonstrate that the morphology of a 
merger at the epoch of its massive dusty starburst depends strongly
on the initial orbital configuration.

The derived difference in global morphology between gas-rich major
mergers with strong dusty starburst provides important implications
 for  the nature of some ULIRGs, though the present results can be
true for mergers without remarkable bulges.
Recent high resolution optical/near-infrared imaging of ULIRGs
have confirmed that morphology of ULIRGs is very diverse and furthermore
that the separation of  two galactic cores is very different ranging
from 0 kpc (single nuclei) to about  $\sim$ 10 kpc (Scoville et al.
2000; Surace et al. 2000).
For example, the core separation
is estimated to be about 25 kpc 
for ULIRG IRAS01199-2307
and just 2.5 kpc for IRAS22491-1808 (Surace et al. 2000).
Since strong dusty starburst occurs in the first encounter
(i.e., when the separation of two disks is an order of 10 kpc)
for the $PP$ model without bulges, it
is not unreasonable to say that such an ULIRG as IRAS01199-2307
is formed by a nearly prograde-prograde merger without remarkable
bulges.
Furthermore, there are several ULIRGs (e.g., IRAS 10173+0828,
VIIZw 031) that do not show obvious
evidence of current or past interactions (Sanders \& Mirabel 1996).
Our results suggests that such ULIRGs are formed by mergers
that do not produce strong and long tidal tails: ULIRGs
without obvious tidal features are formed by nearly retrograde-retrograde
mergers.  We here do not reject the idea that such ULIRGs are formed
by an alternative mechanism (i.e., those other than major merging). 
The present numerical results imply
that the origin of the observed morphological diversity
of ULIRGs is due partly to the difference in orbital configurations
of major mergers.

Fig.5 and Fig.6 show the SED at the epoch
of maximum starburst for 
the two models ($PP$ and $RR$) with and without dust extinction.
By comparing the $UV$ flux with $\lambda < 3000$ $ \rm \AA$
for the model without dust extinction and that for the model with
dust extinction, we can clearly  determine  to what degree
the $UV$ light from secondary starburst is absorbed by
dust.  
Clearly, 
internal dust extinction
of the $RR$ merger is appreciably larger than  that of 
the $PP$ one.
The mean value of $A_{\rm V}$ for all stellar components
is 1.35 for the $PP$ model and 1.50 for the $RR$ one.
Furthermore, 
if we estimate 
the mean value of $A_{\rm V}$ in the central
0.5 kpc of a merger  at the epoch of maximum starburst,
it is  8.06  mag for
the upper galaxy and 6.27 mag for the lower one in the $PP$ model
and 8.42 mag in the $RR$ one.
 There are two main reasons 
for this dependence.
The first reason is  that owing to
a smaller amount of dusty interstellar gas tidally
stripped away from the merger,
a larger amount of interstellar gas can be transferred
to the surroundings of the central compact starburst 
and thereby obscure the starburst more heavily in the $RR$ model. 
The second reason  is  that in the $RR$ model, the  massive starburst
occurs $only$ one-time and
$only$ in the late merger phase when 
the dusty interstellar
gas is the most efficiently transferred 
to the  central region of the merger,
and consequently a larger amount of higher density gas
can obscure the central starburst.
Fig.2 $-$ Fig.6  therefore  imply that both  morphology and SED
in  a dusty starburst merger 
at the epoch of maximum starburst
can be controlled by an
initial orbital configuration of the merger and
thus that the morphology of the merger can 
correlate with the SED.
To be more specific, a dusty starburst merger
without strong and long tidal tails (or with no
clear signs of tidal interaction)
can show very strong internal dust extinction.

Fig.7 and Fig.8  describe how the 
$PP$ and $RR$ merger models 
can be observed by the $HST$ if they are located 
at intermediate and high redshifts ($0.4 \le z \le 1$).
Table 1 furthermore summarizes the apparent $V$ band magnitude,
colours, and sub-millimeter flux at 850 $\mu$m at each redshift for the two
models. 
Most of  the observed ULIRGs are located 
at low redshift (Sanders \& Mirabel 1996),
these figures are accordingly  helpful for deducing morphological
properties of intermediate and high redshift ULIRGs.
Fig.7 suggests that 
the evidence for  major merging becomes less clear
in optical band at
intermediate redshifts  for  both the $PP$ model and the $RR$ one.
This is  essentially because
the  outer 
low surface brightness  tidal feature 
in a merger is hardly detectable in the present optics of the $HST$.
Compared with the $PP$ model with blue compact morphology ($V-I$ = 2.12 
and $m_{\rm I}$ = 20.48 mag),
the $RR$ model shows optically faint and
very diffuse morphology ($V-I$ = 2.28 and $m_{\rm I}$ = 20.70 mag)
owing to the larger
dust extinction.
We here suggest that some dusty starburst
radio sources 
with optically faint morphology 
recently discovered by Richards et al. (1999) in deep 
VLA (Very Large Array) surveys
are likely to be  these $RR$ mergers with the larger dust extinction.
As is shown in Fig.8,
near-infrared NIC2 image of the $RR$ merger at $z$ = 0.4
shows more clearly the sign
of tidal interaction than WFPC2. 
However, the  $RR$ model with larger extinction shows very 
faint and diffuse morphology even in the NIC2
at $z$ = 1 and also has  redder colours ($R-K$ $ \sim $ 5.8 mag and $I-K$
$\sim$ 4.6 mag). 
This result implies that $some$ of the observed high-redshift
EROs with very faint morphology (e.g., EROs observed by Smail et al. 1999) 
are formed by major mergers  with the larger  dust extinction. 
A growing number of observational results on morphology
and spectrophotometric properties of intermediate and high redshift
ULIRGs are now being accumulated (e.g. Tran et al. 1999).
We accordingly  suggest that it is very worthy to
observationally confirm whether an intermediate and high
redshift ULIRG with no clear signs of tidal interaction
preferentially shows the higher degree of internal dust extinction
and redder colours
and is thus a host galaxy for a faint radio source in VLA
surveys and an ERO with optically faint
and very diffuse morphology.

\section{Discussion}

There are several observational results which
imply a physical relationship between morphology
and photometric properties in ULIRGs.
Auri\`ere et al. (1996) found that ULIRGs
with no apparent signs of interaction
preferentially show the larger ratio
of infrared flux to $R$-band one among ULIRGs with
different  morphologies  and thus suggested that these
ULIRGs are
highly obscured by internal dust. 
Arp 220, which is an ongoing merger without strong 
and long tidal tails and an ULIRGs at z$\sim$0.018 (See Fig.1 
in Ohyama et al. 1999 for the clear morphology of  Arp 220),
is observed to show very large internal dust extinction with  
$A_{\rm V}$   $\sim$ 10 estimated from optical
and near-infrared colours (Scoville et al. 1998) and
$A_{\rm V}$ = 45 
from near infrared and mid-infrared emission lines (Genzel et al. 1998). 
The VII Zw031 (IRAS F05081+7936),   
which also appears to have no strong tidal tails in its  R-band image,
is found to suffer very strong UV light extinction 
with the magnitude about 30 times larger than that of other two
ULIRGs with clear signs of tidal interaction 
(Trentham et al. 1999).
Two possible interpretations for these ULIRGs with
the larger  $A_{\rm V}$  are described as follows.
One  
is that these ULIRGs are just in the late merger phase 
when outer tidal tails become very diffuse and thus less remarkable 
owing to the dynamical relaxation during merging: Irrespectively of
merger orbital configurations, mergers can become ULIRGs with very
large $A_{\rm V}$ just in the late merger phase. 
The other is that
the above ULIRGs with very 
large internal dust extinction  can be formed by major mergers
that do not produce any strong tidal tails:
The above ULIRGs are more likely to be formed by nearly 
retrograde-retrograde mergers.
Considering that $outer$ tidal  tail(s) or  features
can be appreciably seen
even in the late phase of galaxy merging
(e.g., NGC3921 in Schweizer 1996),
the latter interpretation seems to be more plausible.
However, since the outer low surface brightness tidal features
are hard to be detected especially for distant mergers and ULIRGs 
(Mihos 1995; Bekki et al. 1999),
finer and deeper  image  data on morphology of  ULIRGs 
are indispensable for determining which interpretation is more 
plausible for the origin of ULIRGs with very large $A_{V}$.

The present results furthermore provide an important clue to the
physical origin for the observed colour differences
between high redshift (z $\ge 1.0$) faint
SCUBA sources. 
Smail et al. (1998, 1999) revealed 
that $V-I$ colour is very different between faint SCUBA sources
and discovered two EROs
with $I-K \ge 6.0$ and 6.8 among  SCUBA sources.
The origin
for  this observed colour differences, which can result from 
several factors such as  differences
in internal dust extinction, redshifts, and the energy
flux ratio of thermal starburst activity to non-thermal AGN
between  SCUBA sources,
has not been clarified at all.
As is described in the Table 1,
our dusty starburst models with 850$\rm \mu m$ flux ranging from 
1.4 to 2.2 mJy
for 1 $\le$ $z$ $\le$ 2  
predict that the $R-K$ ($I-K$) colours at $z$=1, 1.5, and 2 are
5.38 (4.29), 5.96 (4.99), and 6.07 (5.43), 
respectively, for the $PP$ model and  
5.78 (4.57), 6.74 (5.41), and 7.26 (6.19), respectively,   for the $RR$ one.
These results therefore imply that 
$one$ of important factors for the observed colour differences between
 faint SCUBA
sources is the difference in orbital configurations between
higher redshift major galaxy mergers with dusty starburst.
One of observational tests which can assess the validity
of this interpretation is to reveal the physical relationship
between the detailed morphology of SCUBA sources 
and  their colours (or SEDs).
It is, however, considerably  difficult even for the present optics
of the $HST$ to reveal the detailed morphology of high redshift mergers
(Mihos 1995; Bekki et al. 1999).
Thus future large  space 
and ground-based  telescopes  will discover
the fine structure of high redshift SCUBA sources and EROs and
thereby clarify the relative importance of merger orbital configurations
in determining the SEDs and colours in high-redshift dusty starburst galaxies. 

\begin{acknowledgements}
Y.S. thanks to the Japan Society for Promotion of Science (JSPS)
Research Fellowships for Young Scientist.
\end{acknowledgements}

\clearpage


\begin{table}
\caption{The redshift evolution of $V$ band magnitude, colours,
and 850 $\mu$m flux
for the $PP$ model and the $RR$ one. For z = 0 (actually = 0.017
similar to the redshift of Arp 220),  upper and lower values
represent the results of the model without dust extinction and those
of the model with dust extinction, respectively.
 For the models  without dust extinction at z = 0, the absolute magnitude
is given whereas for z $\ge$ 0, the apparent one is given in each of the two
models with dust extinction.}
\begin{center}
\begin{tabular}{lccccccc}
\hline
\hline
$z$  &  $m_V$&  $B-V$&  $V-R$&  $V-I$&  $R-K$&  $I-K$& $S_{\rm 850 \mu m}$ (mJy) \\
\hline
\multicolumn{8}{c}{PP model}\\
\hline
z=0.0& -22.02&  0.245&  0.297&  0.669&  2.171&  1.800& 0 \\
\hline
z=0.0&  14.37&  0.878&  0.593&  1.123&  2.526&  1.996& $5.38 \times 10^2$  \\
z=0.4&  22.59&  1.175&  1.321&  2.116&  3.707&  2.913& 3.07\\
z=1.0&  25.97&  0.696&  0.927&  2.011&  5.376&  4.291& 2.15\\
z=1.5&  27.31&  0.420&  0.637&  1.606&  5.962&  4.992& 1.83\\
z=2.0&  28.07&  0.375&  0.464&  1.124&  6.074&  5.413& 1.58\\
z=3.0&  29.15&  1.334&  0.488&  0.976&  5.919&  5.430& 1.25\\
\hline
\multicolumn{8}{c}{RR model}\\
\hline
z=0.0& -21.94&  0.280&  0.320&  0.698&  2.209&  1.832& 0 \\
\hline
z=0.0&  14.61&  0.953&  0.628&  1.185&  2.674&  2.118& $4.39 \times 10^2$ \\
z=0.4&  22.98&  1.375&  1.432&  2.283&  3.900&  3.049& 2.53\\
z=1.0&  26.82&  1.214&  1.251&  2.469&  5.784&  4.566& 1.80\\
z=1.5&  28.70&  0.629&  1.107&  2.433&  6.737&  5.410& 1.54\\
z=2.0&  29.64&  0.486&  0.675&  1.747&  7.263&  6.190& 1.35\\
z=3.0&  30.86&  1.332&  0.545&  1.139&  7.339&  6.745& 1.08\\
\hline
\end{tabular}
\end{center}
\end{table}


\begin{figure}[htbp]
\epsfile{file=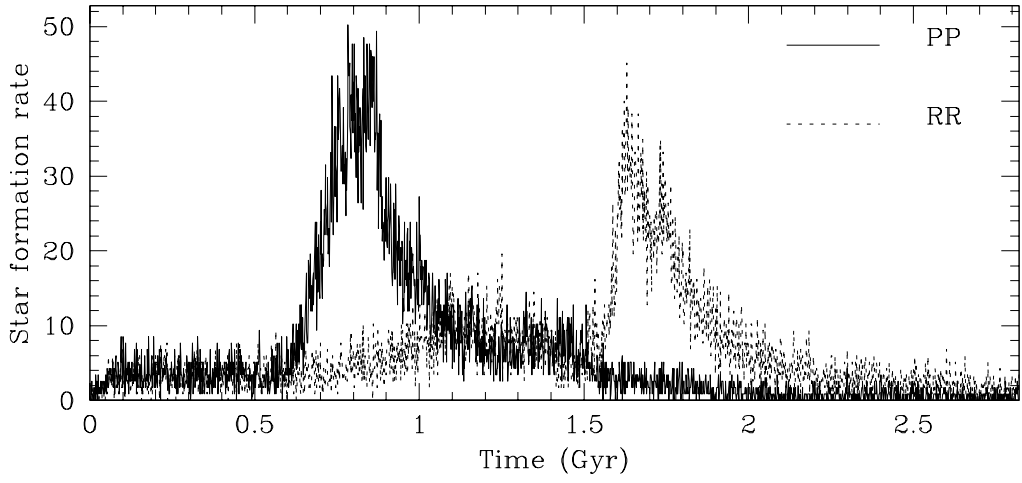}
\caption{
Star formation history for the $PP$ model (solid line)
and the $RR$ one (dotted one).
Star formation is given in units of $M_{\odot}$ ${\rm yr}^{-1}$
(for the model in which the disk mass and size are
6.0 $\times$ $10^{10}$ $ \rm M_{\odot}$
and 17.5 kpc, respectively).
}
\label{fig-1}
\end{figure}

\begin{figure}[htbp]
\epsfile{file=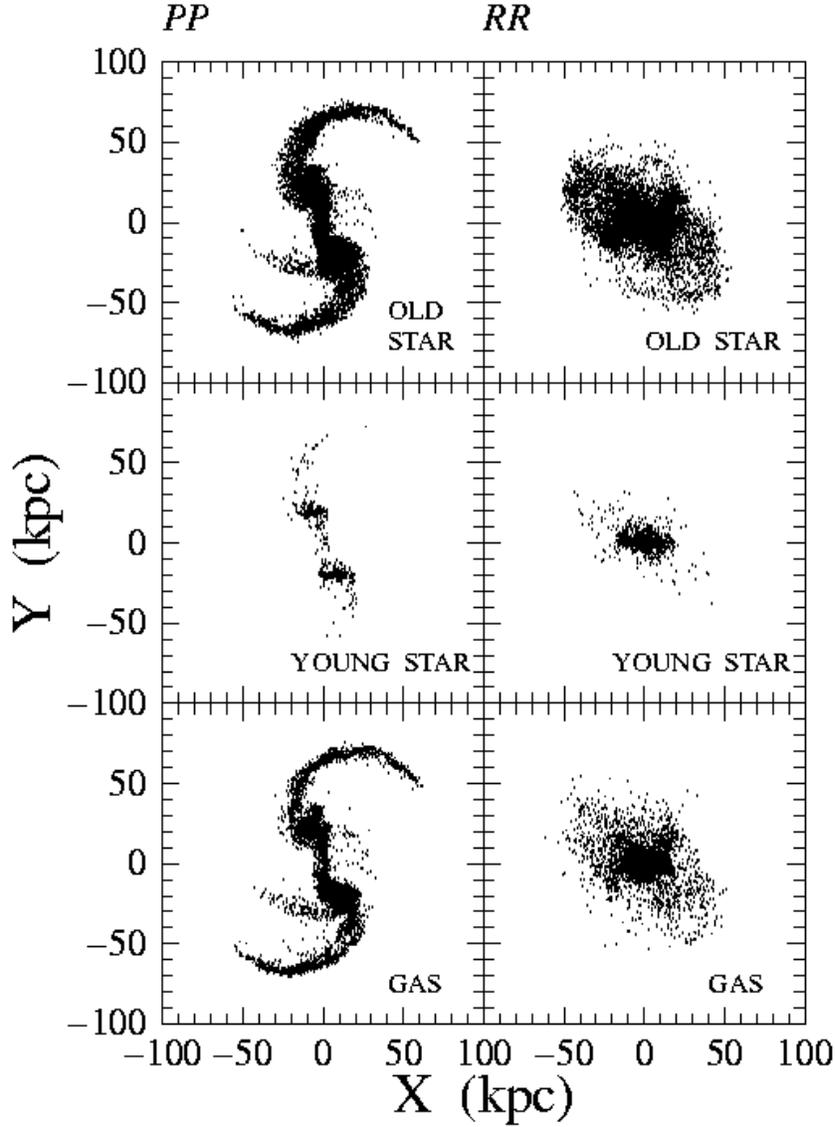,height=17cm}
\caption{
Mass distribution of a galaxy merger projected onto $x$-$y$ plane   at
the epoch of maximum secondary starburst of the merger 
for the $PP$ model (left) and the $RR$ one (right).
Old stars initially located in two disks, young stars formed by
secondary starburst, and gas are plotted in the top panel, the middle
one, and the bottom one, respectively.
Note that strong and long tidal tails can be seen only in the $PP$ model.
Note also that young stars are more compactly distributed within a merger
than old stars. This result suggests that young stars are more heavily
obscured by dusty gas than old stars in a gas-rich merger.
}
\label{fig-2}
\end{figure}

\begin{figure}[htbp]
\epsfile{file=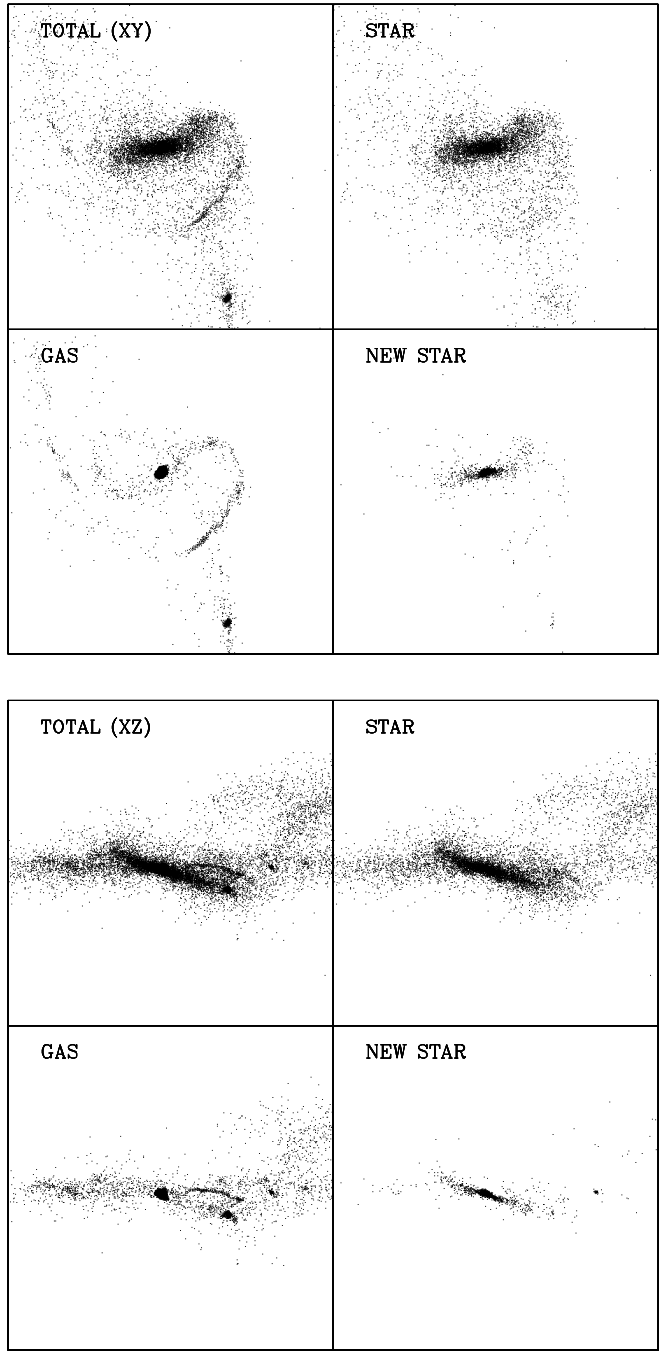}
\caption{
 Mass distribution of the $PP$  model projected
onto $x$-$y$ plane (upper four frames) and $x$-$z$ one (lower four ones)
 at $T$ = 0.7 Gyr corresponding
to the epoch of maximum starburst of the merger for total components 
(upper left), 
old stellar components initially located in two disks (upper right),
gaseous ones (lower left), and new stellar ones formed 
by secondary starburst (lower right).
Each of the eight frames  measures 35 kpc (2.0 in our units)
on a side.
Note that the stellar bar is well developed at
this massive starburst epoch.
}
\label{fig-3}
\end{figure}

\begin{figure}[htbp]
\epsfile{file=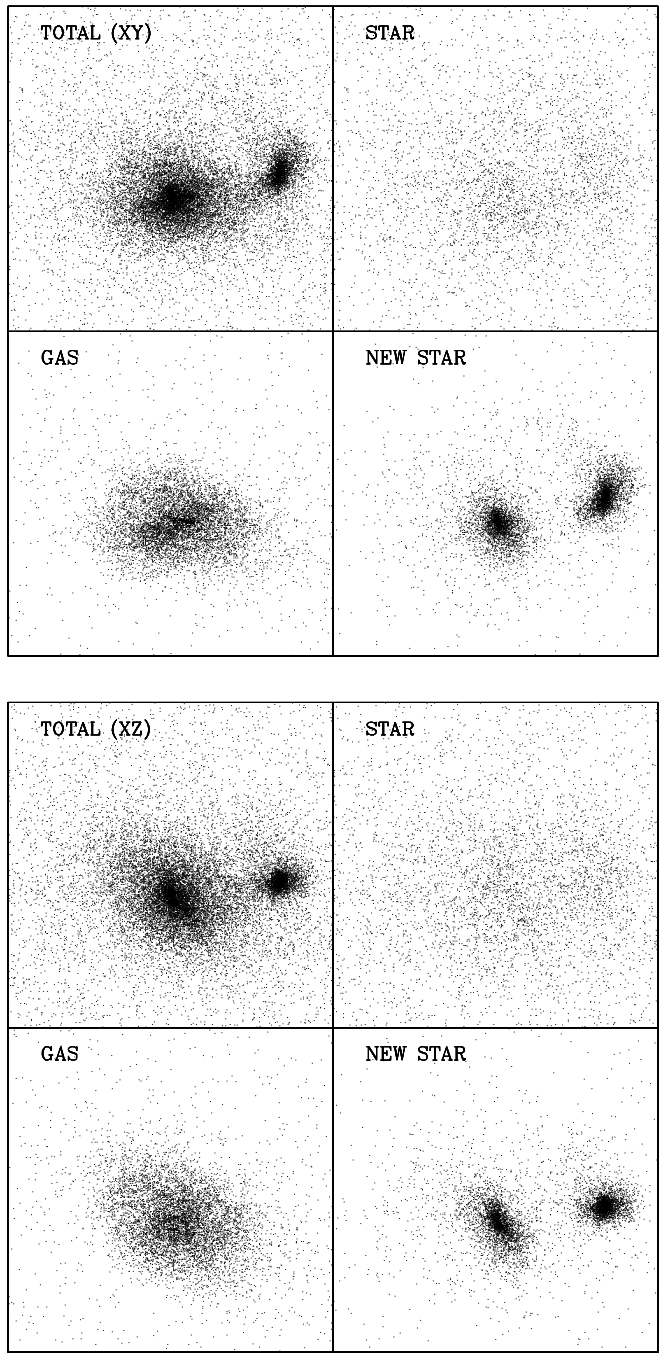}
\caption{
The same as Fig.3 but  for the $RR$ model
at $T$ = 1.6 Gyr. In this figure,
each of the eight frames  measures 3.5  kpc (0.2 in our units)
on a side in order that the central structure
can be more clearly seen for this $RR$ model.
Note that two distinct cores can be seen only in the new stellar
components formed by the star formation in the merger.
}
\label{fig-4}
\end{figure}

\begin{figure}[htbp]
\epsfile{file=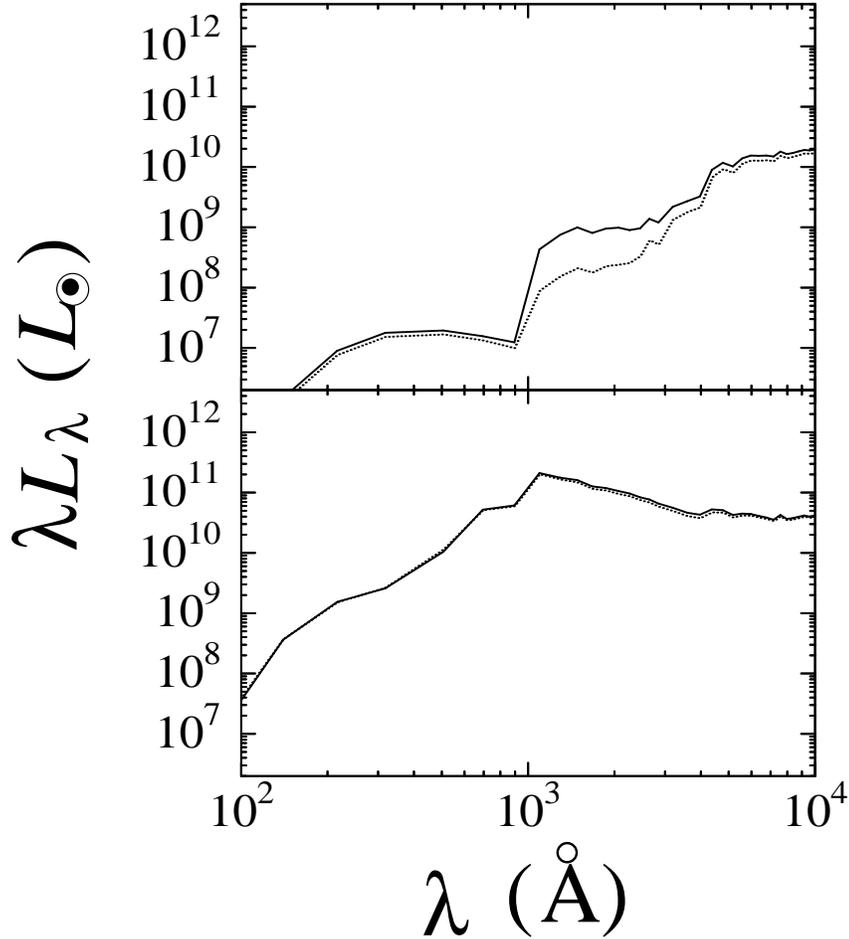}
\caption{
The upper panel shows the  rest-frame SED (with the wavelength
less than $10^4$ \AA)
of a  galaxy merger at the epoch of
its maximum starburst for  $all$ stellar components
in the $PP$ model (solid line),
and for those in the $RR$ one (dotted).
For comparison, the SED of a merger  without dust extinction and re-emission
is also given for each of the two models in the lower panel.
Here  we can clearly 
see the effects of dust extinction and re-emission on the SED shape
in a  merger by comparing the upper panel 
and the lower  one. 
Note that the difference in $UV$ flux  ($\lambda < 3000$ $ \rm \AA$)
between  models with
and without dust extinction is  larger in the $RR$ model.
This indicates that internal dust extinction at the epoch
of maximum starburst (i.e., when a merger becomes an ULIRG)
is the strongest in the $RR$ model.
}
\label{fig-5}
\end{figure}

\begin{figure}[htbp]
\epsfile{file=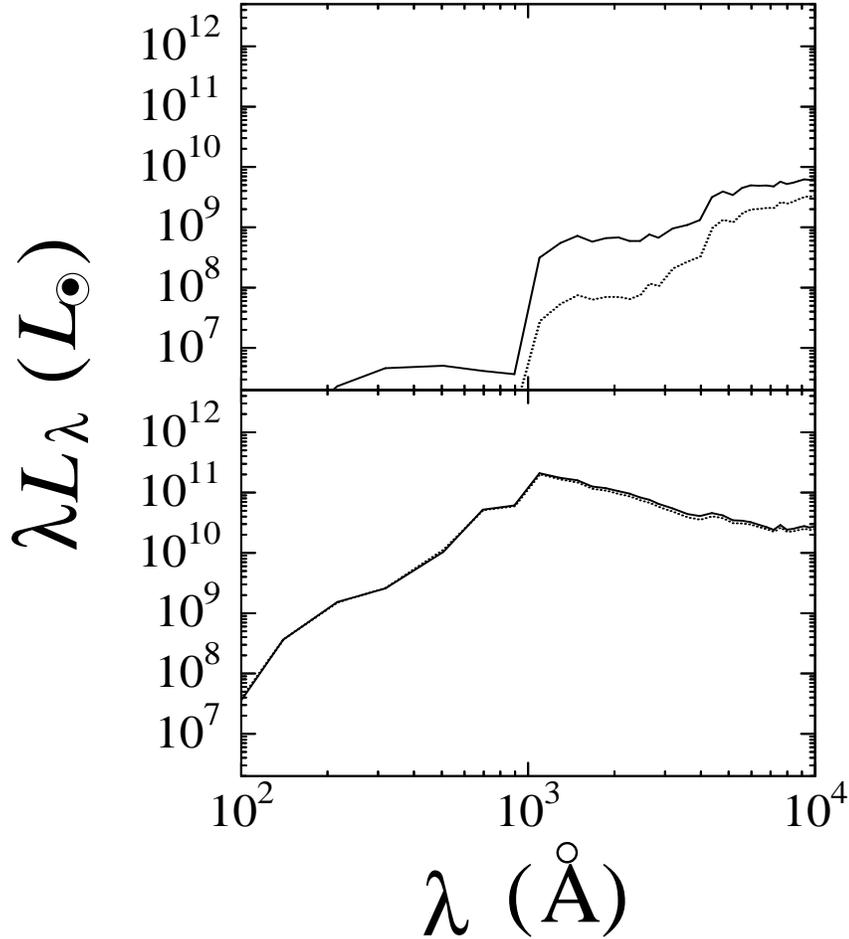}
\caption{
The same as figure 5 but for stellar components within the central
3.5 kpc of the merger in each of the two models.
}
\label{fig-6}
\end{figure}

\begin{figure}[htbp]
\caption{
The $HST$ synthesized image  of  the $PP$
model (upper panel)
and that of the $RR$ one (lower one)
at the redshift z=0.4 
projected onto the $x-y$ plane
corresponding to the orbital plane of the  merger
for the $HST$ WFPC2 (F814W).
In this figure,
 the morphology of the merger  at the epoch of the maximum starburst 
($T=0.7$ Gyr for the $PP$ model and $T$ = 1.6 Gyr for the $RR$ one)
is described.
The method to create this synthesized image is described
in detail by Bekki et al. (1999).
The exposure time for each  synthesized morphology 
is set to be ${10}^{4}$ sec (2000sec $\times5$)  for
the $HST$ WFPC2.
Each frame measures ${9.}^{\prime\prime}$9 
corresponding to 63.7 kpc on a side and
one pixel size is ${0.}^{\prime\prime}$1 for the WFPC2.
}
\label{fig-7}
\end{figure}

\begin{figure}[htbp]
\caption{
The $HST$ synthesized image  of  the $RR$
model at the redshift z=0.4 (upper) 
and 1.0 (lower) for the $HST$ NIC2 (F160W).
In this figure,
the morphology of the merger  at the epoch of the maximum starburst 
($T=1.6$ Gyr) is described.
The exposure time for each  synthesized morphology 
is the same as that of the WFPC2  (${10}^{4}$ sec) in Fig.3.
Each frame measures ${9.}^{\prime\prime}$9 
(corresponding to 63.7 kpc) on a side and
one pixel size is 
${0.}^{\prime\prime}$076  for the NIC2. 
}
\label{fig-8}
\end{figure}

\end{document}